\documentclass[journal=jacsat,manuscript=article]{achemso}

\usepackage[version=3]{mhchem} 
\usepackage[T1]{fontenc}       
\usepackage{graphicx,color}

\newcommand{\EF}{$E_{\rm F}$}
\newcommand{\vF}{$v_{\rm F}$}

\newcommand{\STO}{SrTiO$_3$}

\author{Hyejin Ryu}
\affiliation{Advanced Light Source, Lawrence Berkeley National Laboratory, Berkeley, CA 94720, USA}
\alsoaffiliation{Max Planck-POSTECH/Hsinchu Center for Complex Phase Materials. Max Plank POSTECH/Korea Research Initiative (MPK), Gyeongbuk 37673, South Korea}
\altaffiliation{These authors contributed equally to this work.}

\author{Jinwoong Hwang}
\affiliation{Department of Physics, Pusan National University, Busan 46241, South Korea}
\altaffiliation{These authors contributed equally to this work.}

\author{Debin Wang}
\affiliation{The Molecular Foundry, Lawrence Berkley National Laboratory, Berkeley, CA 94720, USA}

\author{Ankit S. Disa}
\affiliation{Department of Applied Physics and Center for Interface Structures and Phenomena, Yale University, New Haven, CT 06520, USA}

\author{Jonathan Denlinger}
\affiliation{Advanced Light Source, Lawrence Berkeley National Laboratory, Berkeley, CA 94720, USA}

\author{Yuegang Zhang}
\affiliation{The Molecular Foundry, Lawrence Berkley National Laboratory, Berkeley, CA 94720, USA}
\alsoaffiliation{Physics Department, Tsinghua University, Beijing 1000864, China}

\author{Sung-Kwan Mo}
\affiliation{Advanced Light Source, Lawrence Berkeley National Laboratory, Berkeley, CA 94720, USA}
\email{SKMo@lbl.gov}

\author{Choongyu Hwang}
\affiliation{Department of Physics, Pusan National University, Busan 46241, South Korea}
\email{ckhwang@pusan.ac.kr}

\author{Alessandra Lanzara}
\affiliation{Materials Sciences Division, Lawrence Berkeley National Laboratory, Berkeley, California 94720, USA}
\alsoaffiliation{Department of Physics, University of California, Berkeley, CA 94720, USA}

\title
  {Temperature-dependent electron-electron interaction in graphene on \STO}


\begin{document}

\begin{abstract}
The electron band structure of graphene on \STO\ substrate has been investigated as a function of temperature. The high-resolution angle-resolved photoemission study reveals that the spectral width at Fermi energy and the Fermi velocity of graphene on \STO\ are comparable to those of graphene on a BN substrate. Near the charge neutrality, the energy-momentum dispersion of graphene exhibits a strong deviation from the well-known linearity, which is magnified as temperature decreases. Such modification resembles the characteristics of enhanced electron-electron interaction. Our results not only suggest that \STO\ can be a plausible candidate as a substrate material for applications in graphene-based electronics, but also provide a possible route towards the realization of a new type of strongly correlated electron phases in the prototypical two-dimensional system via the manipulation of temperature and a proper choice of dielectric substrates. 


\end{abstract}

The interaction between two-dimensional (2D) materials and three-dimensional (3D) substrates not only induces novel physical properties in their interfaces~\cite{WangE, HeS, OhtomoA} but also provides a viable route towards the comprehension and manipulation of the characteristics of 2D materials themselves~\cite{HwangC}. For example, the dielectric response of the substrates strongly affects the electron-electron interaction in graphene, resulting in a renormalized Fermi velocity (\vF) and reshaped Dirac cone~\cite{BarlasY,HwangEH,EliasDC,ChaeJ,SiegelDA}. Of particular interest, \vF\ near the Dirac point increases with enhanced electron-electron interaction, defying the ordinary Fermi liquid behavior where \vF\ decreases with stronger interaction~\cite{HwangC}. Finding a suitable substrate to morph the properties of 2D materials into a desired way is a key to improve the efficiency and functionality of 2D material based devices~\cite{GeimAK}.

\STO\ is an ideal candidate as a substrate with complex layers of interesting physical properties. Its dielectric constant varies by a couple of orders of magnitude with varying temperature ($T$)~\cite{WeaverHE} and it exhibits quantum paraelectricity at low temperatures~\cite{RowleySE}. More interesting properties emerge when \STO\ is interfaced with other materials, such as increased superconducting phase transition temperature of single-layer FeSe above 100~K~\cite{Ge}, two-dimensional electron gas~\cite{Meevasana, Santander-SyroAF}, superconductivity~\cite{ReyrenN, UenoK}, extremely high mobility~\cite{SonJ}, electronic phase separation~\cite{Ariando}, and the coexistence of magnetism and superconductivity~\cite{LiL}. Although the microscopic mechanism of such emerging phases is still a matter of intense investigation, the list of novel properties emerging from various materials interfaced with \STO\ is ever expanding.

Considering the current interest in both 2D materials and \STO, it is surprising that studies of graphene on \STO\ are limited to a handful of transport data~\cite{Morpurgo,Sarma,SahaS, Sachs} in the literature. A recent transport study~\cite{SahaS} reports anomalous `slope-break' in resistivity with temperature range of 50 K - 100 K and increasing mobility at low temperatures, which is not observed in graphene/SiO$_2$. The origin of such unconventional transport properties has been attributed to the structural phase transition of the \STO\ substrate~\cite{SahaS}. On the other hand, another study shows that the transport properties of graphene are unaffected by the \STO\ substrate when the external magnetic field is zero~\cite{Morpurgo}. Instead, thermal conductivity measurement on graphene shows an anomalous enhancement and subsequent breakdown of Wiedemann-Franz (WF) law within the similar temperature range, suggesting that the intrinsic electronic correlations in graphene are highly dependent on temperature as a consequence of the formation of strongly interacting Dirac electrons, so-called Dirac fluid~\cite{Kim}. Whether a similar temperature dependence can be found in spectroscopic measurements and how it may manifest itself in the measured spectra are essential information required to further understand such anomalous transport data. 

In this Letter, we report a $T$-dependent angle-resolved photoemission (ARPES) study of graphene on an \STO\ substrate. The graphene sample was prepared by the chemical vapor deposition method~\cite{Hong2009} using methane and transferred onto the \STO(001) substrate after annealing the substrate under O$_2$ flow to remove any residual oxygen vacancies (see Methods for detailed sample preparations). The graphene on an \STO\ substrate is then annealed in ultra high vacuum to remove contaminants for ARPES measurements. At $T\sim180~{\rm K}$, the deviation from the characteristic linearity in the measured energy-momentum dispersions mostly follows the conventional picture~\cite{HwangC} of dielectric-assisted enhancement of electronic correlations albeit the dielectric constant of an \STO\ surface is heavily renormalized compared to that in bulk. Such deviation is enhanced at low temperatures, which cannot be explained within the dielectric picture in multiple ways detailed below. We also report that the spectral width at Fermi energy (\EF) and \vF\ of graphene on \STO\ are comparable to those of graphene on a BN substrate, suggesting that \STO\ can also be a plausible candidate as a substrate material for applications in graphene-based electronics.


The graphene samples on the \STO(001) substrate were characterized by both Raman and ARPES. Figure~1(a) shows Raman spectra taken for several different samples. The blue curve (G/STO) is a raw Raman spectrum taken from graphene on \STO. In order to extract signal from graphene, a Raman spectrum for a bare \STO\ substrate (red curve: STO) was subtracted from the G/STO spectrum resulting in the black curve (G/STO$-$STO) that shows peaks at 1580~cm$^{-1}$ and 2700~cm$^{-1}$. These peaks correspond to the primary in-plane vibrational mode, G band, and a second-order overtone of a different in-plane vibration, 2D band, of graphene, respectively~\cite{Ferrari}. The black curve agrees well with the typical Raman data taken from graphene on SiO$_2$, which is the green curve (G/SiO$_2$). One can also notice that graphene on \STO\ shows negligible signal corresponding to the disorder-induced D band around 1350~cm$^{-1}$~\cite{Ferrari}. The inset shows optical microscope images of graphene on \STO\ and SiO$_2$ for comparison. 

Figure~1(b) shows constant energy intensity maps from ARPES measured on the same sample. ARPES is a sensitive tool to characterize the quality of graphene by directly probing its electron band structure, e.\,g.\,, charge neutrality~\cite{Sprinkle}, band curvature~\cite{SiegelDA}, number of layers~\cite{OhtaPRL}, and effect on chemical/dielectric environment~\cite{HwangC}. For example, one can decide the charge neutrality from the position of the Dirac point, and layer numbers from how many linear bands exist in the electron band dispersion. The ARPES also provides the information on the quality of the graphene from the linewidth of the spectra~\cite{SiegelNP}. Fermi surface ($E-E_{\rm F}=0.0~{\rm eV}$) consists of several spots that expand to wider crescent-like shape at higher energies, each of which shows the characteristic conical dispersion of charge neutral single-layer graphene. There exist three major domains of graphene with different azimuthal orientations as indicated by blue, green, and gray hexagons, within the probing area for ARPES measurements of $\sim$40$\times$80~$\mu$m$^2$ (photon beam spot size of the measurements), consistent with the azimuthal disorder of typical CVD-grown graphene~\cite{Incze}.

  \begin{figure*}
  \begin{center}
  \includegraphics[width=0.5\columnwidth]{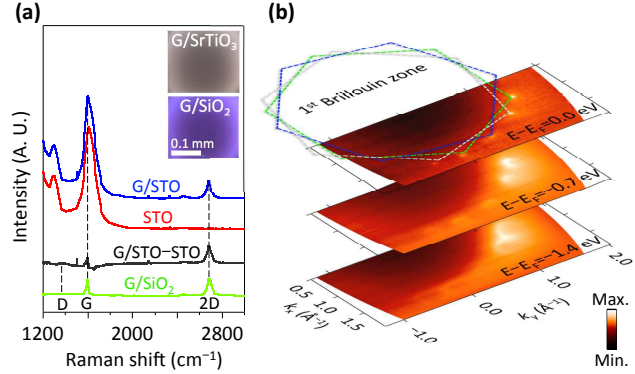}
  \end{center}
  \caption{(a) The blue curve (G/STO) is a raw Raman spectrum taken from graphene on \STO. The black curve (G/STO$-$STO) is obtained by subtracting a reference spectrum taken from a bare \STO\ substrate (red curve: STO) from the G/STO spectrum to extract the contribution from graphene alone. The green curve (G/SiO$_2$) is a spectrum taken from graphene on SiO$_{2}$ for comparison. Insets are optical microscopy images of graphene on \STO\ (up) and SiO$_{2}$ (down). (b) Constant-energy ARPES intensity maps taken at $E-E_{\rm F}=0~{\rm eV}, -0.7~{\rm eV}, {\rm and} -1.4~{\rm eV}$.}
  \label{Fig1}
  \end{figure*}

Figure~2(a) shows an ARPES intensity map taken along the $\Gamma-{\rm K}$ direction of hexagonal Brillouin zone (BZ) as denoted by the white line in the inset. Along this orientation, only one of the two branches of the graphene $\pi$ band is observed due to the matrix element effect~\cite{HwangQP}. The single linear band crossing $E_{\rm F}$ evidences the single-layer nature of the graphene on \STO~\cite{OhtaPRL}. The red curve in Fig.~2(b) is the momentum distribution curve (MDC) taken at $E_{\rm F}$ from the data shown in Fig.~2(a). The full width at half maximum, $\Delta\, k$, of the MDC for graphene on \STO\  (red curve) is 0.037~{\rm \AA}$^{-1}$, even narrower than that for graphene on BN (blue curve), 0.044~{\rm \AA}$^{-1}$~\cite{HwangC}. The BN substrate is considered as one of the ideal substrate materials for graphene inducing an order of magnitude higher charge carrier mobility~\cite{DeanCR} and significantly improved  temperature and electric-field performance compared to graphene on SiO$_{2}$~\cite{MericI,ScheirtzF}. This is due to the flat nature of graphene on BN, whose roughness is three times less than that on a SiO$_{2}$ substrate~\cite{DeanCR}. From an ARPES point of view, the flatness of graphene and the enhanced mobility are observed as a narrow spectral width at $E_{\rm F}$~\cite{HwangC}, where self-energy contribution is almost vanished, and as an enhanced \vF\ near \EF, respectively. The latter can be directly extracted from Lorentzian fits to the MDCs at each energy of the graphene $\pi$ band, which is shown in Fig.~3(a). The observed \vF\ is $1.42\times10^6~{\rm m/s}$, which is significantly enhanced compared to $0.85\times10^6~{\rm m/s}$ within the local density approximation (LDA), but similar to $1.49\times10^6~{\rm m/s}$ measured from graphene on BN~\cite{HwangC}. The sharp spectral width and the high \vF\ comparable to those for graphene on BN suggest that \STO\ can also be one of the ideal candidates as a substrate material for graphene-based electronic devices.

  \begin{figure*}
  \begin{center}
  \includegraphics[width=0.5\columnwidth]{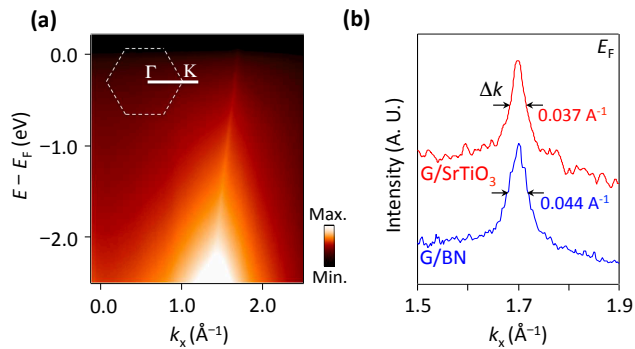}
  \end{center}
  \caption{(a) ARPES intensity map of graphene on \STO\ taken at 13~K along the $\Gamma-{\rm K}$ direction as denoted by the white line in the inset. The white dashed line is the Brillouin zone of the hexagonal unit cell of graphene. (b) Momentum distribution curves at $E_{\rm F}$ for graphene on \STO\ (red curve) and graphene on hexagonal BN (blue curve) taken at 13~K and 15~K, respectively. $\Delta\,k$ denotes the full width at half maximum of the peak as denoted by black arrows.} 
  \label{Fig2}
  \end{figure*}

More interesting features can be observed from the details of the extracted dispersion in Fig.~3(a) and its temperature dependence. First of all, the energy-momentum dispersion at 180~K deviates from the linearity of Dirac electrons as shown in Fig.~3(a). The difference between the measured dispersion, $E({\bf k})$, and the dispersion from LDA, $E_{\rm LDA}({\bf k})$, gives a good approximation of an electron self-energy. Figure~3(b) shows the self-energy as a function of wave number that is well fitted by a logarithmic function, $\Sigma(\mathbf{k})=\alpha\hbar v_{0}/4 \times (\mathbf{k}-\mathbf{k}_{\rm F})\ln(\mathbf{k}_{\rm C}/(\mathbf{k}-\mathbf{k}_{\rm F}))$~\cite{GonzalezJ,SiegelDA}, where $\alpha$ is a fine-structure constant of $e^{2}/(4\pi\varepsilon\hbar v_{0})$, $v_{0}$ is the bare velocity of electrons of $0.85\times10^6~{\rm m/s}$, $\mathbf{k}_{\rm C}$ is the momentum cut off of 1.7~\AA$^{-1}$, and $\mathbf{k}_{\rm F}$ is the Fermi wave number. The vanishing density of states at \EF\ of charge neutral graphene leads to the absence of metallic screening. With the unscreened Coulomb interaction, graphene exhibits the singularity in the interaction and resultant logarithmic correction to the electron self-energy analogous to a marginal Fermi liquid~\cite{GonzalezJ}. Indeed, the logarithmic fit gives $\alpha=1.1$, implying that this system requires a full theoretical treatment beyond the random-phase approximation~\cite{Kotov}. Hence the logarithmic self-energy is a good evidence of non-Fermi liquid behavior of charge neutral graphene~\cite{GonzalezPRL,SiegelDA}. In addition, the logarithmic correction, i.\,e.\,, the curvature of the dispersion, is further enhanced by increasing the strength of electronic correlations~\cite{HwangC}.

  \begin{figure*}
  \begin{center}
  \includegraphics[width=0.5\columnwidth]{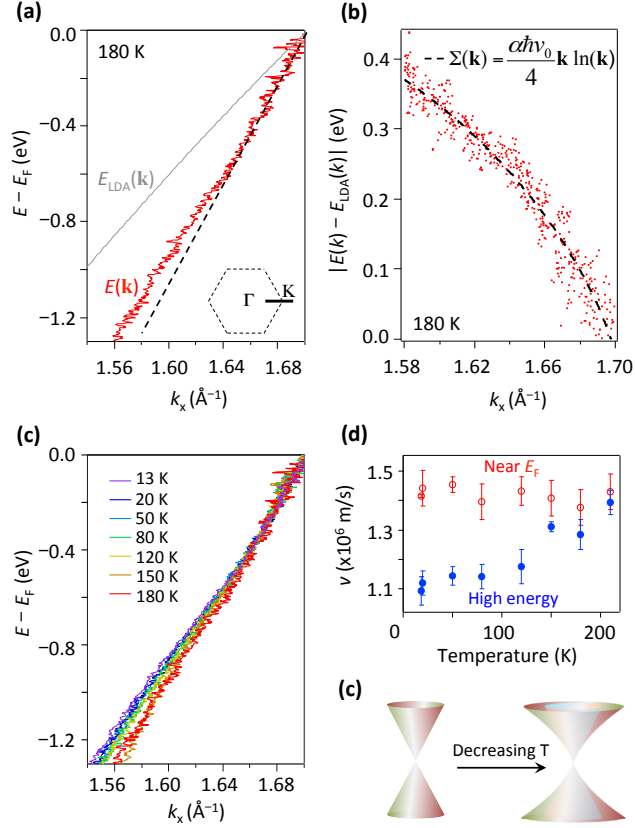}
  \end{center}
  \caption{(a) An energy-momentum dispersion of graphene on \STO, $E({\bf k})$, taken along the $\Gamma-{\rm K}$ direction at 180~K. The gray solid line, $E_{\rm LDA}({\bf k})$, is an LDA band and the black dashed line is a straight line extension from the lowest energy part of the dispersion, for comparison. (b) An $|\,E({\bf k})$-$E_{\rm LDA}({\bf k})\,|$ dispersion at 180~K. The black dashed curve is a logarithmic fit to the $|\,E({\bf k})$-$E_{\rm LDA}({\bf k})\,|$ dispersion. (c) Energy-momentum dispersions of graphene on \STO\ taken along the $\Gamma-{\rm K}$ direction at several different temperatures. (d) The slope of the dispersion obtained by a line fit to the dispersion between $0.0~{\rm eV}$ and $-0.4~{\rm eV}$ (red empty circles) and between $-0.6~{\rm eV}$ and $-1.1~{\rm eV}$ (blue filled circles) as a function of temperature. (e) Cartoon illustrating the electron band structure of graphene near charge neutrality at high (left) and low (right) temperatures. }
  \label{Fig3}
  \end{figure*}

The effective dielectric constant of graphene can be extracted from $\alpha$~\cite{HwangC,SiegelDA}. For the data taken at 180~K, extracted dielectric constant of graphene is only 2.33, which is comparable to that of suspended graphene, 2.2$\sim$5~\cite{EliasDC}, but completely different from the dielectric constant of bulk \STO\ (300$\sim$18000)~\cite{WeaverHE}. The apparent discrepancy suggests that the dielectric screening from the \STO\ substrate to the graphene is overwhelmingly suppressed. The formation of an interfacial layer between a metal and an insulator often leads to a few orders of magnitude reduced capacitance than the bulk value of the insulator~\cite{WeaverHE,Mead}, which is known to be an intrinsic property unavoidable at the interface due to the rearrangement of atoms to compensate strains~\cite{Stenger,Chang}. Such a dear layer has been also reported at the surface of a dielectric film such as \STO~\cite{CZhou}. The formation of a dead layer on the surface of \STO\ from annealing process strongly reduces the dielectric constant from 400 (bulk) to 140 (film) at 180~K and from 18000 (bulk) to 1000 (film) at 2~K~\cite{WeaverHE,BasceriC}. In our case, the TiO$_2$-terminated surface was exposed not only to the air, that contaminates a surface with hydroxides, hydrogen molecules, etc., but also to chemicals used during the transfer of graphene on top of \STO, which will result in an ill-defined crystalline structure of the surface of \STO\ (see Supplementary Information I for detailed discussion). This will lead to the change of the dielectric property of \STO. Indeed, the dielectric constant of \STO\ very close to the surface can be as low as $\varepsilon\sim5$~\cite{Kiat}. This implies an effective dielectric constant at the \STO\ surface of $\sim$3 using the simple standard approximation, $\varepsilon=(\varepsilon_{\rm vacuum}+\varepsilon_{\rm surface})/2$, similar to the extracted dielectric constant for graphene on \STO\ from our ARPES data. Thus both the interfacial effect between graphene and \STO, and the ill-defined surface of \STO\ are attributed to play an important role in the dramatic change of the dielectric property of the \STO\ surface. Our result is consistent with the transport data showing that the dielectric constant of \STO\ does not affect the transport properties of overlying graphene~\cite{Morpurgo}. The depressed dielectric screening on the surface of \STO\ leads graphene to retain its intrinsic properties such as charge neutrality and resultant strong electron-electron interaction despite graphene stands on a substrate.

The strongly suppressed dielectric constant and the charge neutrality of graphene on \STO\ allow graphene to revive its intrinsic strength of electronic correlations despite graphene is placed on a substrate. As discussed in Fig.~3, the logarithmic correction to the electron self-energy is a good evidence of non-Fermi liquid behavior of charge neutral graphene~\cite{GonzalezPRL,SiegelDA} and it is further enhanced with increasing strength of electronic correlations~\cite{HwangC}. As shown in Fig.~3(c), the energy-momentum dispersion of the graphene $\pi$ band is further modulated upon decreasing temperature. The slope at higher energy gradually decreases, whereas the slope near Fermi energy barely changes as summarized in Fig.~3(d) and schematically drawn in Fig.~3(e). In other words, the curvature of the energy spectrum becomes even stronger with decreasing temperature than the logarithmic correction can describe (see Supplementary Information II for detailed discussion).

The accelerated enhancement of electron-electron correlation at low temperatures is similar to the recent experimental report on the strongly coupled Dirac fermions in charge neutral graphene~\cite{Kim}. With decreasing temperature, thermal conductivity of graphene exhibits an abnormal upturn, which evidences the breakdown of WF law due to the formation of Dirac fluid states near charge neutral point. The upturn of the thermal conductivity was found to be $\sim$120~K, similar to the temperature where the slope at higher energy starts to decrease in our measurements. While the lower limit of the formation of Dirac fluid is the temperature corresponding to the disorder potential in the presence of charge puddles for the case of graphene on SiO$_2$ ($T_{\rm dis}\sim40~{\rm K}$)~\cite{Kim}, the sharper spectral feature discussed in Fig.~2 in conjunction with the negligible D band discussed in Fig.~1(a) might allow us to speculate much smaller $T_{\rm dis}$ in graphene on \STO, so that the Dirac fluid state persists down to the experimental low temperature limit in our measurements. Although our results do not directly prove the formation of the Dirac fluid states in graphene on \STO, both studies suggest that temperature is an important factor driving strong electron-electron correlation in charge neutral graphene.


In conclusion, we have reported the spectroscopic evidence of temperature-dependent non-linearity of the energy spectrum of graphene on \STO. The sharp spectral width and the high \vF\ comparable to those for graphene on BN suggest that \STO\ is one of the plausible candidates as a substrate material for graphene. More importantly, strongly suppressed dielectric screening on the surface of \STO\ allows us to manipulate the electron self-energy of graphene as a function of temperature. The modified electron self-energy is attributed to the enhanced electron-electron correlation at low temperature, which cannot simply be described by a dielectric screening picture. Our results suggest that temperature in conjunction with a proper choice of a substrate can lead to a strong electron-electron correlation in graphene. Making use of the enhanced electronic correlation of graphene despite the existence of a substrate will provide a versatile platform for the artificial modification of a functionality of graphene-based devices.

{\bf Experimental Section.} The TiO$_2$-terminated SrTiO$_3$(001) crystal was obtained from CrysTec GmbH. To remove any residual oxygen vacancies, the crystal was annealed under O$_2$ flow for 6 hours at 600~$^{\circ}$C~\cite{Tan}. The quality of \STO\ was confirmed by atomic force microscopy (AFM) and four-point resistivity measurements. AFM shows unit cell high steps (~4~\AA) and surface roughness below 1/2 unit cell (1.9~\AA) (see Supplementary Information III). The resistivity of >10~M$\Omega$ confirms the removal of dopants. The graphene sample was grown on a copper foil by chemical vapor deposition (CVD~\cite{Hong2009}) using methane at 1000~$^{\circ}$C. The samples were placed in a tube furnace (Lindberg Blue) and pumped to $\sim$100-500 mTorr. 35 sccm of H$_2$ (99.999\%) was flowed during the heating to the CVD temperature, 1000~$^{\circ}$C. The growth was done by flowing a mixture of H$_2$ (2 sccm) and CH$_4$ (35 sccm) for 2 hours. The system was cooled down under a flow of 35 sccm H$_2$. A 500~nm thick layer of polymethylmethacrylate (PMMA) was spin-coated on top of the graphene, followed by the removal of copper by wet etching ($\sim$30\% FeCl$_3$ and $\sim$4\% HCl in deionized water). The graphene sample was then transferred onto the SrTiO$_3$(001) substrate, followed by annealing at 680~$^{\circ}$C in ultra high vacuum with a base pressure of $1\times10^{-10}$~Torr. ARPES measurements were performed at the Beamlines 4.0.3 and 10.0.1 of the Advanced Light Source, Lawrence Berkeley National Laboratory, using photons with energies of 90~eV and 50~eV, respectively. The energy and angular resolutions set to be 18~meV and 0.2~$^\circ$, respectively. Raman measurements were performed using a laser source with a wavelength of 532~nm.

{\bf Supporting Information} The electron band structure of graphene/\STO, the electron self-energy of graphene/\STO\ as a function of temperature, an AFM image of \STO.

\begin{acknowledgement}
We would like to thank J. Yi for helpful discussions. This work was supported by Berkeley Lab's program on sp2 bond materials, funded by the U.S. Department of Energy, Office of Science, Office of Basic Energy Sciences, Materials Sciences and Engineering Division, of the U.S. Department of Energy (DOE) under Contract No.~DE-AC02-05CH11231. The work in Max Planck POSTECH Center for Complex Phase Materials was supported by the National Research Foundation of Korea (NRF) funded by the Ministry of Science, ICT and Future Planning (No.~2016K1A4A4A01922028). The work in Pusan National University was supported by Basic Science Research Program through the National Research Foundation of Korea (NRF) funded by the Ministry of Science, ICT and Future Planning (No.~2015R1C1A1A01053065 and No.~2017K1A3A7A09016384). The Advanced Light Source is supported by the Office of Basic Energy Sciences of the U.S. Department of Energy under Contract No. DE-AC02-05CH11231. 
\end{acknowledgement}

\end{document}